# Supporting MOOC instruction with social network analysis

Tanmay Sinha
Department of Computer Science
Vellore Institute of Technology
Chennai 600127, India
tanmay.sinha655@gmail.com

## ABSTRACT
With an expansive and ubiquitously available gold mine of educational data, Massive Open Online courses (MOOCs) have become the an important foci of learning analytics research. In this paper, we investigate potential reasons as to why are these digitalized learning repositories being plagued with huge attrition rates. We analyze an ongoing online course offered in Coursera using a social network perspective, with an objective to identify students who are actively participating in course discussions and those who are potentially at a risk of dropping off. We additionally perform extensive forum analysis to visualize student's posting patterns longitudinally. Our results provide insights that can assist educational designers in establishing a pedagogical basis for decision-making while designing MOOCs. We infer prominent characteristics about the participation patterns of distinct groups of students in the networked learning community, and effectively discover important discussion threads. These methods can, despite the otherwise prohibitive number of students involved, allow an instructor to leverage forum behavior to identify opportunities for support.

## Categories and Subject Descriptors
K.3.1 [**Computers and Education**]: Computer Uses in Education; K.3.1 [**Computers and Education**]: Distance learning - *Massive Open Online Course, Learner Engagement Pattern*

## General Terms
Algorithms, Experimentation, Human Factors

## Keywords
Coursera, MOOC, Social network analysis

## 1. INTRODUCTION
With the evolution of a multitude of content delivery options and learning paradigms, Massive Open Online Courses (MOOCs) have profoundly changed the way we think of education in terms of both practical education, life-long learning outside the confines of an institution and academic achievements. Online education at such a massive scale has the potential to be very transformative, because it provides learners with openness and autonomy about when and how to learn, what to learn and how to construct and apply the knowledge that comes of it. The rationale behind the design of these MOOCs is the underlying theory of connectivism [1], which stresses more on interaction with other participants and on the student-information relationship. Because of the option of free and open registration, a publicly shared curriculum and open ended outcomes, very often it has been observed that there is massive enrollment in these digitalized MOOC courses.

However, as McAulay et al. [2] elucidates, participation in a MOOC is "emergent, fragmented, diffuse, and diverse". Sharp drop off rates characterize most of the online courses from content providers like Coursera, Udacity, EdX etc [3]. From genuine sources such as MIT technology review, official mailing lists of the tutor, it can be inferred that this pattern of decreasing participation is persistent across beginner level courses such as "Circuits and Electronics" (offered by EdX, Spring 2012, completion rate 0.045) to advanced level courses like "Quantum Mechanics and Quantum Computation" (offered by Coursera, July 2012, completion rate 0.014). Therefore, gaining insight into the communication dynamics of these MOOCs is an interesting research direction, that can provide answers to the indispensable question of why there are large drop outs in MOOCs. Because MOOCs make a powerful promise of providing students with an unprecedented level of global access to vast set of educational opportunities, such answers are critical for the development of effective computer mediated support for students.

In our work, we choose a social network analysis (SNA) lens to analyze the MOOC data. SNA helps us to decipher informal connections between interacting participants, which may not otherwise be visible. It enables us to view where are students embedded within the network, their structural influence and importance as discussion initiators or responders. Apart from finding out the social structure hidden behind the MOOC network that facilitates knowledge sharing, we also incorporate detailed forum analysis of students' posting patterns changing over time to augment our understanding of different categories of students' conversations. The objective is to automatically highlight potential problems for the instructors beforehand, about specific groups of students whose participation patterns are uncommon. For example, by having the prior knowledge of students who are exhibiting signs of dropping off from the course, instructors can target these groups of students, thereby providing them specialized support based on learning style, available time and personal objectives. On the other hand, a prior indication about students who are showing signs of active engagement in course discussions, can enable instructors to provide them with apt incentives for continuing to share their knowledge.

This paper is organized as follows: In Section 2, we review prior work relevant to this domain. Section 3 describes the course and context of our study. In Section 4, we describe our approach followed for social network formation. In Section 5, we characterize the MOOC network with both macro (groupings and relationships) and micro (individual perspective) analysis. This helps us to further dive into the analyzing the discussion forums of the chosen MOOC in detail in Section 6. We summarize major findings of our work and discuss future avenues of exploration in Section 7 and Section 8, and conclude in Section 9.

## 2. RELATED WORK
Traditional social network analysis (SNA) has been used for a long time to study online communities like Internet relay chats [4][5][6][7][8], health communities [9], popular question answer forums like Java [10], Yahoo [11], Naver [12] etc. with multi-ended objectives. These objectives include expertise detection, monitoring knowledge generation and user participation, predicting best answers in question answer forums by combining user's social network and answer characteristics, exploring learning outcomes and group interaction processes. From the context of computer supported collaborative learning (CSCL), since Haythornthwaite et al. [13] formally defined a generic approach to examine e-learning by social network methods, interaction patterns within networked learning communities have been investigated using this approach thereafter [14][15]. Because of the absence of a direct connection between users in multiparticipant conversation analysis, effective methods are required to represent the structure of networks resulting from the interaction of participants. Therefore, approaches like temporal coherence, reply structure, conversation proximity and word context have been the primary methods for social network formation to throw light on the structural dynamics of communities.

Prior work on preventing university learning and distance education from getting plagued by dropouts, has mainly dealt with comparison of machine learning techniques and trying to identify the most appropriate learning algorithm [29][30][31]. Cheng et al. [22] developed two scalable tools - micro commitment button and an action plan, to "identify students at the risk of dropping off from the class". Deployment of these tools in a Coursera HCI course yielded positive results in detecting and improving students' engagement. The diverse degrees of participation as defined by Fischer et al. [23] and further substantiated by Dick et al. [24], were incorporated into two MOOC courses designed on the OpenHPI platform [25]. This case study qualitatively showed that active course participants achieve higher scores and are responsible for the social quality of the learning context. Belanger et al. [26] described the development and delivery of "Bioelectricity" course offered on Coursera in 2013, analyzing students' pre-course motivations for enrolling and contrasting the participation trajectories of students who earned a certificate with those who didn't. Major barriers that emerged out of statistical and general conversation analysis included lack of time, insufficient background skills and difficulty in applying concepts learnt. Similarly, Schimdt et al. [27] discuss the lessons learned after the offering a Coursera course on "Pattern Oriented Software Architecture" in 2013. The instructors stressed on challenges such as time constraints involved in content preparation and managing MOOCs after the start, need for constant monitoring of discussion forums to accelerate learning processes, removing common misconceptions, building good will and rewarding positive participation. Despite some innovations devised for this course such as virtual office hours, crowd-sourced assignments and a grade calculator, a low completion rate of 0.051 was achieved, similar to prior MOOC offerings.

This reinforces the fact that there is something fundamentally wrong in the way MOOCs aim to foster global life-long learning communities by connecting a motley of students. There is a missing link that prevents instructors and course designers to understand students' behavior fully. Therefore, in our work, we apply social network methods that adds richness to MOOC analysis and enables us to discover interesting facts about the student participation that can't be effectively discovered only by statistical approaches. Our analysis also augments the understanding of discussion forums and students' characteristics informally exhibited in the MOOC, thereby potentially enabling instructors to filter out important students or interesting threads of discussions to focus on.

## 3. CONTEXT

For our purpose of study, we chose a course offered through Coursera. Having started in April 2012 with just 30 online courses, Coursera has now developed into a full fledged online education portal featuring 395 courses from 84 renowned Universities extending beyond one discipline, and 9.5 million enrollments from students representing 195 countries.

We selected a single active course to study. Intuitively, we wanted to identify an ongoing course at an introductory level (expertise or prior knowledge was not required to join), whose course content was more generic (not limited to a specific discipline) and where there was scope for more discussion and knowledge building with the subject matter. Therefore, we selected "Fantasy and Science Fiction: The Human Mind, Our Modern World" (offered in June 2013) as our preliminary course for analysis.

We collected 1503 posts and 1100 comments among 665 threads posted to the discussion forums during the first seven weeks of the course. A total of 771 students participated in the forum discussion during the seven-week period, not counting those students posting anonymously.

Generally, Coursera discussion portals are divided into forums and subforums and each of them have certain number of threads. Each thread is initiated with a "thread starter" post that serves as a prompt for discussion. The thread builds up as people start follow up discussions by their posts and comments. Our data contains all posts and comments for the given period, annotated with thread, forum, subforum, author (when not posted anonymously), parent post (for comments), and timestamp.

## 4. EXTRACTING A SOCIAL NETWORK

Because Coursera has a structured discussion forum, we used the reply structure to form our social network for the "Fantasy and Science Fiction: The Human Mind, Our Modern World" MOOC. "Thread Starters," or discussion initiators, have an outward link to all people within a particular thread. If people posted more than once within a thread, they would have a stronger tie strength to the thread starter. Intuitively, this meant that they were more connected to the thread starter. Focusing on the directionality of links was essential here, because an undirected (symmetric) network would lessen importance of these discussion initiators. However, we also noticed that there were people within threads, who engaged people in active discussions by their posts. In other words, their posts within a thread generated lots of comments. Thus, we could informally identify formation of sub threads within each thread. So, we labeled people as "Sub-thread starters", if their posts generated more than 3 (chosen arbitrarily) comments within a thread. Following a similar approach as thread starters, these sub thread starters were connected by outward links to people who commented on their posts within a thread.

To form our final social network graph, we took a union of the above 2 networks, so that we could emphasize the connections of both thread starters and sub thread starters. We did so by matching node labels and summing duplicate ties. This gave a network that reflected the true social structure of the MOOC, laying adequate importance on majority of the discussion initiators. We collected data from the discussion forum twice. The first collection occurred after the course had been running for 3 weeks. The constructed social network consisted of 397 nodes (students) and 1024 edges (interactions among students). Then, we scraped the data again after the end of the 7th week to see how it had changed over time. The seven-week network consisted of 771 people and 3848 edges.

# 5. CHARACTERIZING THE MOOC NETWORK

In this section, we describe characteristics of the "Fantasy and Science Fiction: The Human Mind, Our Modern World" MOOC network. This will serve as a base to compare and contrast our findings with the detailed forum analysis, as described in Section 6. It will also help us to figure out how significant is the network structure for understanding students' posting patterns in the course forum. The social network perspective emphasizes multiple levels of analysis. However, we consider the macro and micro views of the network respectively. The extent to which individuals are connected to others (micro view) and the extent to which the network as a whole is integrated (macro view), are two sides of the same coin. Differences among individuals in how connected they are can be extremely consequential for understanding their attributes and behavior. More connections often mean that individuals are exposed to more diverse information. Highly connected individuals may be more influential, and may be more influenced by others. On the other hand, more connected populations may be better able to mobilize their resources, and bring multiple and diverse perspectives to bear to solve problems.

## 5.1 Macro View
The macro view of the network enables us to focus on groupings, relationships and the roles of users interacting in the network.

### 5.1.1 Bow Tie Network Analysis
The Bow/Tie structure is one of the fundamental ways of observing the network skeleton. Originally proposed for visualizing the structure of the World Wide Web (WWW) [16], the structure contains 6 main components: In, Out, Scc (Strongly connected component), Tubes, Tendrils and Others (disconnected network nodes). There is a directed path from nodes of 'In' component to all the nodes of 'Scc' and a directed path from nodes in the 'Scc' to every node in 'Out' component. Also, hanging off 'In' and 'Out' structures are 'Tendrils', which contain nodes that are reachable from portions of 'In', or that can reach portions of 'Out', without passage through 'Scc'. It is possible for a 'Tendril' hanging off from 'In' to be hooked into a 'Tendril' leading into 'Out', forming a 'Tube', that is a passage from a portion of 'In' to a portion of 'Out' without touching 'Scc'.

In the MOOC data, Figure 1 depicts the Bow Tie network structure for four different combinations as specified by the rows and columns, each component showing the proportion of students belonging to it. Thread starters will primarily belong to the 'In' component. Among them, few students who can be reached from one another using directed links, form the 'Scc' component. 'Out' component represents students who reply in threads/sub threads created by students in 'Scc'. 'Tubes' represent those students who connect students in 'In' component to the 'Out' component without passing through the 'Scc'. Important inferences that we can make from the Bow Tie structure analysis are:

- Observing vertically, the proportion of increase in people from the 'Scc' to 'Out' component is almost nil in the final scrape. This indicates that the group of students who are strongly connected amongst themselves are not drawing in participation from other students later in time. Or, we can hypothesize that this network structure provides some evidence that, people who join later are not much integrated into the ongoing active discussions. Furthermore, the fact that proportion of people in the 'Scc' component is same, indicates that only very few people are actively engaging in discussions persistently.

- Observing horizontally, the proportion of people in 'Tubes' component has increased, after emphasizing relationships of sub thread starters in the network. This means that the increase corresponds to these sub thread starters, who only act as bridges between thread starters and other students. It is because the proportion of people in 'Scc' remains same. So, we can deduce the fact that the sub thread starters are not engaging in active discussions among themselves.

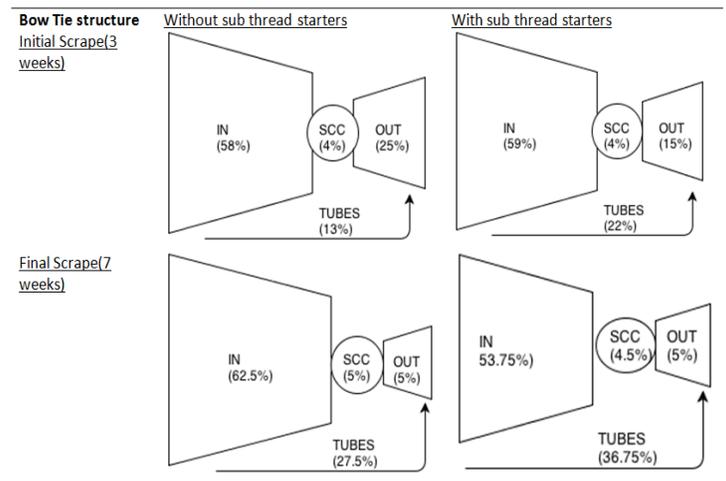

**Figure 1. Bow-Tie network structure for the MOOC**

## 5.2 Micro View
The micro view of the social network enables us to have an individualistic view and helps us in selecting individuals, whose interaction patterns seem interesting. By having this prior knowledge, we can a) generalize the social network properties, and, b) study changes in students' social network metric over time, to map it to changes in posting behavior or linguistic patterns in their conversations.

### 5.2.1 Centrality
In social networks, the motivation for defining centrality measures arises, because certain structural locations in the network are more advantageous or disadvantageous to nodes than others. Therefore, centrality measures help us in understanding the notion of power attached to these "highly favored positions" embedded in the relational network, in terms of opportunities offered and constraints imposed. For our purpose of analysis, we considered the basic measures of degree and betweenness centrality. Degree centrality for a particular node is defined as the addition of the number of inlinks and outlinks. On the other hand, betweenness centrality of node 'x' is basically a measure that sums up the proportion of shortest paths between all node (i,j) pairs in the network on which node 'x' sits on. These centrality measures are primarily linked with how important a node is in the network. For example, people with high betweenness centrality act as bridges between two different groups of people. So they are influential in spreading across ideas of one group to the other. Figure 2 depicts the sociogram for the given MOOC under study, with nodes colored by betweenness centrality and sized by out-degree centrality. This means larger sized nodes would have higher degree centrality, while darker colored nodes would have higher betweenness centrality.

Because the Coursera discussion forum allowed anonymous postings and it was not possible to point out specific students who had made anonymous postings within the forums, we removed the "anonymous" node while performing our analysis. From the sociogram, we can observe that there are many students who have high degree centrality as well as proportionately high betweenness centrality. Moreover, we hypothesize that there is a notion of the presence of small and dense components throughout the network. There seems to be different subgroup formation within the network, potentially because of thread starters and sub thread starters. The presence of 58 weakly connected components in this social network corroborates this. Nevertheless, in the next section, we look at the same network structure, but from a different perspective which brings further clarity.

### 5.2.2 Coreness and Centrality Variations

Extending the notion of centrality, it is often observed that social networks exhibit a core-periphery structure. The core is a dominating set that can also be seen as a group with maximum group centrality. As Borgatti et al. [17] points out, a key characteristic of core/periphery structures is that they have relatively short path distances between pairs of nodes, which allows information to flow rapidly. In addition, the core nodes serve as gatekeepers for flows between peripheral nodes. Hence, in a social network, core/periphery structures are ideal for spreading ideas and practices that the core deems are worthy while preventing the spread of unauthorized ideas. The extent to which a node belongs to the graph's core can be thought of as the coreness of the node.

We were interested in validating our hypothesis of section 5.2.1, that a)whether the network complies with the core/periphery model and is there a single big core with a periphery of people around, or b)whether there are many small and dense cores throughout the network. However, to see how coreness of students in the MOOC varies with centrality, we first needed to understand that all coreness measures are centrality measures, but the converse is not necessarily true. For example, it is possible to collect a set of the 'n' most central actors in a network, according to some measure of centrality say, closeness or degree, and yet find that the subgraph induced by the set contains no ties whatsoever (an empty core).

So, this is exactly what happens in our MOOC data. We observe peculiar spikes in variations of students' coreness with decreasing betweenness centrality. This indicates that even with decreasing centrality, many people have high coreness value. This is because such students are connected to different cohesive regions of the graph and do not necessarily have any ties to each other. Even if we think intuitively, in our MOOC data, subgroup formation primarily takes place because of thread starters and sub thread starters who initiate discussion among groups of people who comment on their created threads. They form small subgroups in the graph that are well intra knot, but not closely inter knit.

### 5.2.3 Hypertext Induced Topic Selection (HITS) Score

Coreness and centrality measures mostly consider the count of connections to estimate the prominence of a node. However, they do not characterize and capture the notion of the quality of connections. So, to get an insight into this slightly changed dynamics, we also analyze the HITS score for each node. This is basically a ranking algorithm and assigns 2 scores to each node: one is the Authority score (how valuable information stored at that node is) and a Hub score (quality of the nodes' links) [18]. In our MOOC network, people with a good authority score are those students who engage other students in discussions (how valuable their piece of conversation is). Students with good hub score are those are who get engaged in discussions initiated by many active users (thread starters/sub thread starters). From the HITS authority and hub scores for all people, we observe:

- Students having very high authority scores have very high hub scores too (Category 1 MOOC users). Such students are actively engaged in posting on threads (discussions) of other similar users. This indicates "the rich get richer" social phenomena or their preferential attachment [19] with students who are having greater authority scores.

- Therefore, we get another evidence that supports our hypothesis in Section 5.1.1 of some students, not being drawn into participating in discussions. Such students are the ones who have low hub scores (Category 2 MOOC users). There might be possible explanations: a)One of the reasons might be the fact that they have low authority scores, that is, the piece of information that they bring into the network is not so novel and valuable. b)Another reason maybe the preferential attachment of Category 1 students with similar students, and hence lesser interaction with the Category 2 students.

- From a different viewpoint of the same scores, the other major reason for concern is that people with low authority scores do not effectively engage in active discussions to improve their learning. This is evident from downward spikes in their hub scores.

## 6. FORUM ANALYSIS OF THE MOOC

Social network analysis addresses how ties are patterned among the interacting participants in the MOOC and how these patterns influence ongoing bonds. It gives a high level view of the structural dynamics of the network. We found out that the analysis and evolution of this social structure itself, gives us interesting results for interpretation. However, the Coursera infrastructure that is built around this social network provides unprecedented opportunities for effective knowledge discovery about the inner forum operations. So, in this section, we aim to scrutinize the forums for the "Fantasy and Science Fiction: The Human Mind, Our Modern World" MOOC. This will give us deeper insights into a) the way students are conversing in these forums, or what are their posting patterns and, b)whether the forum structure can act as a lens to view the nature of conversations. Similar to other Coursera MOOCs, this online course is divided into 6 forums- General discussion, Frequently asked questions, Assignments, Technical issues, Study groups and Unit discussion. Furthermore, the Unit discussion forum is subdivided into 10 subforums from Unit 1 to 10. All forums and subforums have different threads of activity, uniquely identified by their thread number

### 6.1 Evolution of Discussion forums

As a first step, to get a glimpse of overall forum activity, we plot the posts and comments of people using our final data scraped up till 7 weeks. We notice that while study groups draw initial interest, participation in them decreases immensely after week 4. Such study groups are meant to connect people having similar demographics such as common work time (schedule), background, location etc. However, the discussion in these study groups becomes sparse as time progresses. This may be explained by: a)Limited knowledge of students, which is insufficient to spark off interesting discussions and, b)the social phenomenon of Homophily [20]. Though students in study groups are "birds of the same feather", they may have novelty seeking tendencies. That is, they would not want to meet a student similar to them in every aspect, because it will provide no new information/stimulation (boredom trap). The likelihood of connection will drop off automatically, if they don't find interesting topics for conversation and reasons for being engaged (curiosity). Therefore, the only remaining social force that can drive them to participate is curiosity.

We also observe that Unit Discussion and Assignment forums have uniformly dense postings throughout. So, we need to further zoom into these forums to see what threads are interesting here, what threads are being returned back to at a later point in time and which posts bring back discussions in these old threads. Therefore, as a logical next step, we look at what conversation categories help to grow these discussion forums. In the given MOOC, we qualitatively observe the "Unit Discussion" forum and notice that when students usually return back to post back in old threads, there are there intense pockets of discussion. To confirm this quantitatively, we cluster threads in all these forums based on factors like thread length (number of posts within a thread), content length (number of characters within a thread), thread duration (difference between the timestamp values of the first and last thread), thread density (thread length/thread duration) and content density (content length/thread length).

The intuitive idea behind clustering is that it will help to distinguish between these different conversation categories simultaneously going on in the MOOC - 'short and detailed' (discussions on specific, rare and advanced topics or discussions of expert people in which not many people will be drawn into) versus 'long and detailed' (healthy on topic discussions) versus 'short and less detailed' (uninteresting threads or new threads which have just started) versus 'long and less detailed' (factual discussions or introductory social conversations) versus 'bursty (intense) pockets of discussion.

Figure 2 shows a multi-attribute visualization of the results of K Means clustering (with random initialization of cluster points, scoring metric as "distance to centroids", distance measure as "Euclidean" and number of clusters optimized from 2 to 6). This visual representation along with the quantitative values of each thread, gives us the following inferences:

- Cluster C5 represents threads that are intense pockets of discussions and are not detailed, but consist of only short responses (very high thread density). On further exploration, we see that majority of posts in these threads are spread between General discussions, Assignments and Unit discussion forums.

- Cluster C2 represents threads that are intense pockets of discussions which are detailed too (very high content density).

- For cluster C1, we see that it has the maximum thread duration, very less thread density, but comparatively higher content density. This means that when threads return back, there are very few postings, but these are more detailed.

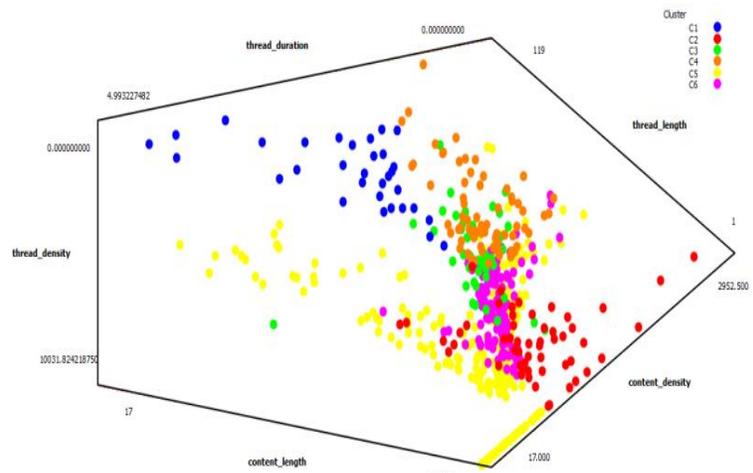

**Figure 2. Clustering threads based on thread length, duration, density; content length, density**

Some examples of interesting threads are shown in Table 1.

## 6.2 The "Unit Discussion" forum

Instead of broadly studying all forums in this MOOC, we narrow down our focus to the Unit discussion forum. A zoomed in analysis of the posting patterns of students as they progress through units and their posting patterns as the essay (assignment) deadline approaches, would be interesting to discover how do students engage with each unit. The "Fantasy and Science Fiction: The Human Mind, Our Modern World" MOOC is structured in a way that that each unit is taught in one week and the unit's videos release date and assignments due date fall two days before the week's end.

After visualizing the forum behavior, important inferences that we can elicit are:

- Till Unit 4, people's posts and comments usually start towards the end of the week (because unit videos are released 2 days before the end of the week) and it continues roughly till 1.5 weeks later. There is not much discussion before the assignment due dates.

- Unit 5 and Unit 6 are exhibit an interesting posting pattern. There is some initial interest a week before these units actually start, but there are very few posts and comments in the allotted weeks.

- Another fascinating thing that we observe is that, there are few outlying posts that come "early" with regard to the unit. These outliers basically consists of posts and comments from students who have some pre knowledge about the unit or are familiar with the science fiction work. Either they know about the text of the book or things related with the unit like e-books, audio books, movie releases of the same fiction work etc. Some examples of such posts are shown in the Table 2.

**Table 1. Thread categories discovered from clustering**

| Category | Characteristic | Text |
|---|---|---|
| 1)Limited Knowledge of people in | High duration, Very less thread length | *"Are there any other scientists out there interested in forming a* |

| | | study group?.....I'm a Wildlife Ecologist and Science Teacher.... Although not a huge Fantasy fan myself, I have been enjoying the use of animal imagery and symbolism in Grimm's Fairy Tales.....I find it really entertaining to point out the scientific mistakes the authors made in their work, such as blood transfusion in Dracula" |
|---|---|---|
| Study groups | | |
| 2)Homophily in study groups | Very low content density, Very high thread density, Very low duration | Consists of mainly Introductions |
| 3)Opening up new ideas | High content density (belongs to cluster C2) | " So, I missed my " assignment due" Date... BLERGH! On the bright side, that gives me an opportunity to post my thoughts here, and I'm not limited to a measly 320 words....... I come at this from the viewpoint of Female, Anachronistic.....I'm unconvinced that there was a deliberate concealment of pre-Christian beliefs in the tales....... In light of some of the info in Prof Rabkin's video on the methodology and motivations of the Grimms in gathering these tales......" |
| 4)Posting one's own perceptions | Very less Thread duration with Content density much greater than Thread density (belongs to cluster C1) | " I could feel the 1950's in this book and can't help but feel we have made progress as a society. Not that we won't want parades and I suspect we will act in the same fashion....." |

**Table 2. Outlying posts**

| Post | Text |
|---|---|
| 1) Post 354- Unit 4 | " I know that it's best to read slowly and make annotations while reading, BUT, I came across this **free audio download** of Frankenstein......" |
| 2) Post 2704- Unit 4 | " Those of you who were lucky enough to see the **National Theatre's 2011 production of "Frankenstein"**-either live in London, or via one of the worldwide cinema broadcasts......." |
| 3) Post 2695- Unit 9 | " Should anyone want to borrow and then delete my version which is **online-readable** , please contact..." |
| 4) Post 2325- Unit 7 | " From a slumber as deep as death, as refreshing as that of a healthy child....... This is the first sentence in the third chapter of "Herland"..... **This sentence makes that kind of sense of wonder I love in science fiction and fantasy....**" |
| 5) Post 1552- Unit 8 | "I could feel the 1950's in this book and can't help but feel we have made progress as a society...... **Lovely read**" |

## 6.3 Important Groups of students

In sections 6.1 and 6.2, we analyzed the posting patterns for every student, in all the MOOC forums and also in the individual units of the Unit discussion forum. However, to co-relate our results of social network analysis with forum analysis, it would be more apt to look at specific groups of students and observe their longitudinal posting pattern. So, we consider the following two groups of students as described:

### 6.3.1 Cohorts

Cohorts are basically those groups of people who start the course within the same week. So, after identifying cohorts by their first post week, the intriguing research questions was: what forums were these cohorts interested in posting, that is, their overall posting patterns with time. This would help us to know how did these cohorts engage with the course material, and which units were the later cohorts posting or commenting in. Figure 3 tells us that cohort 1 and 2 are persistently engaged in unit discussions for all units, though the persistence decreases after each unit. The X axis represents the duration in weeks from 1 to 7, while the Y axis represents the unit number from 1-10 in the course. The colored dots represented in the figure corresponds to different cohorts from 1 to 7. These figures provide an evidence that people who join early persist longer. We also find that majority of the early posts in the later units, as found out in Section 6.2, are made mostly by cohort 1 and 2 only.

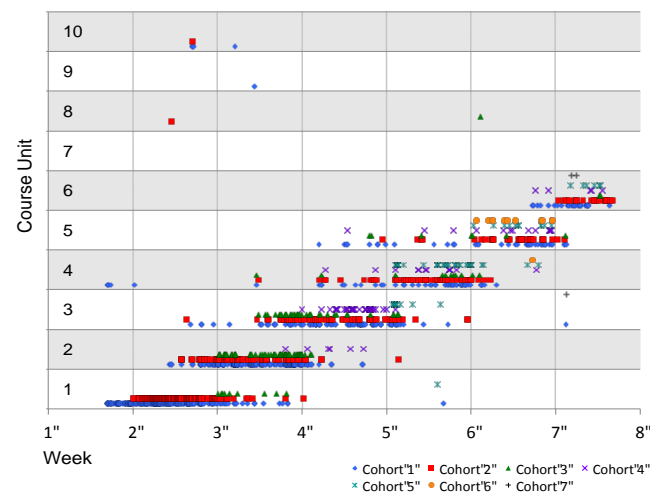

**Figure 3. Posting pattern of cohorts per week per unit**

Another important observation is that later cohorts do not start participating from the beginning. Their posts and comments start only from one or two units before their joining week. Cohorts 2, 3 and 4 exhibit an interesting pattern. Their number of posts is greatest in the unit that was just taught before their week of joining the course. This is indicated by the spikes for these cohorts in Figure 4. Here, X axis represents Units from 1 to 10, while the Y axis represents number of posts. Each colored line represents the participation trajectory of cohorts from 1 to 7.

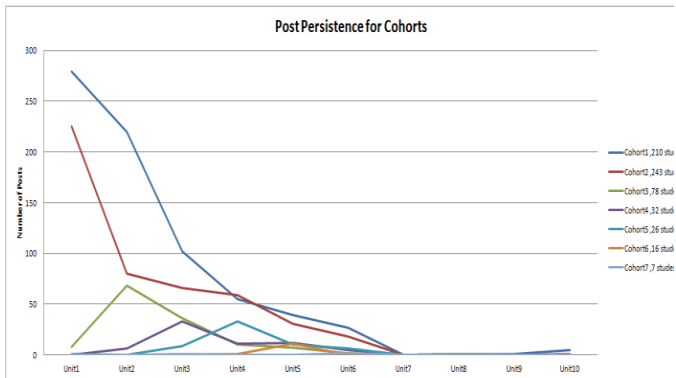

**Figure 4. Post persistence for cohorts**

### 6.3.2 Students with decreasing centrality

The other group of students we looked at were basically those students who remained central over time (as inferred by the centrality social network metric) and those who didn't. The objective was to figure out what trend in their posting patterns made them more central or decreased their centrality over time. For our analysis, we looked at people who remained within the top 20 degree centrality list and those who did not, from the initial scrape of the MOOC data (3 weeks) to the final scrape (7 weeks). Analyzing more deeply, we noticed that for students who remained central over time, half of them were engaging in dense discussions though all the units in the Unit discussions forums. The other half of these students actually retained their centrality by their large number of postings on other course forums as opposed to Unit discussions forum. We observed that students who didn't remain central over time had almost nil posts and comments in the Unit discussions forum. An important finding here that correlates with our earlier observations is that most of these students who became central over time, belonged to earlier cohorts like 1 and 2 only. Also, a nice characteristic of these people is that they exhibit good behavior and etiquette in the forums. Some examples are shown in the Table 3.

**Table 3. Good behavior from students in earlier cohorts**

| Post | Characteristic | Text |
|------|---------------|------|
| 1)Post 4375 | Warning students of bad behavior in forums | "Everyone should take Javier with a pinch of salt. I've seen him around a couple of my other courses and he just gets off on trolling and trying to rile people up. Just ignore his obnoxious comments. " |
| 2)Post 3437 | Motivating others | "Megan explains things brilliantly as always, I second every word. I can relate to feeling you have more interesting ideas you'd like to share. Remember, you can always post them on the forum. While you won't get grades, you may get to discuss your thoughts in more detail than in the essay, and get different opinions that will widen your perception of the book ....." |
| 3)Post 818 | Calming down students after a heated argument | "Thank you all for your input. I really did not mean to spark a heated debate, and for that, I am sorry. Frankly, I am surprised at how many people are okay with this system..... I always thought I was a progressive educator, but there are some things in which I will not budge. At this point, we must agree to disagree. I am glad I made someone laugh at least, if nothing else. Happy reading all!" |

To gain further understanding into the communication dynamics of these central and non central students over time, we also looked at their ego networks. As Welser et al. [21] pointed out, analysis of ego networks can help us differentiate "discussion persons" from "answer persons" in an online discussion. Answer people tend to reply to discussion threads initiated by others and typically only contribute one or two messages per thread. They are disproportionately connected to alters with low degree, that is, they are tied to many alters who themselves have few ties and few alters who have many ties. The answer person's network is primarily sparse, star-shaped, and has numerous inward connections from relative isolates. On the other hand, discussion persons are those who contribute initial turns that elicit brief replies or who typically reply to threads initiated by others with large numbers of additional messages. The discussion person's network has dense ties to highly connected alters. When discussion people are tied to relative isolates, those ties are often directed outward, indicating that the relative isolates contact the discussion person.

So, looking at the ego networks of students who remained within the top 20 degree centrality list over time in the MOOC, we found out that : a) Majority of the students who remained central, had started off as being answer persons, but most of them turned to discussion persons as time progressed, b) On the contrary, majority of the students who didn't remain central, started off as being discussion persons, but most of them turned to answer persons as time progressed. This re-emphasized the importance of discussions in MOOCs and its attached relevance to students' importance.

## 7. DISCUSSION AND SUMMARY

In this section, we summarize our major findings, implications of our research and insights on what we have learnt from MOOC analysis. We categorize the summary into 3 logically coherent and connected parts: Students, MOOC Discussions and Social network structure

## 7.1 Students

Students who engage in active discussions form a very small subset of MOOC participants. At the beginning of the course, they do communicate with new students. However, as the course moves on, these students are engrossed in talking only among themselves. They do not effectively engage other students of the course in discussions. As found out by Social network metrics, there are the following possible reasons for this: a) The piece of information that other students bring into the network is not so novel and valuable; b)"The rich get richer" social phenomena or preferential attachment amongst the actively participating people; c)A third reason can be the fact that, the purpose for students to join the MOOC may not be solely discussion oriented, that is, students may have other objectives like just viewing the lectures, only reading discussion posts etc., rather than actually engaging in discussions. Thread starters form a major proportion of these students. On the other hand, we also observed that students who are discussion initiators within threads, only act as bridges to connect thread starters with other students in the MOOC network. In other words, these sub thread starters do not engage themselves in active discussions.

Next, students who start in earlier weeks (Week 1 and Week 2) have persistent posts across Unit discussions forum, that is, such people persist longer. However, the persistence decreases with time. These students remain central as time progresses, and make outlying posts early with regard to the unit. These posts and comments indicate some pre knowledge about the unit, familiarity with the science fiction work and knowledge about things related with the unit like e-books, audio books, movie releases of the same fiction work etc. On closer inspection, we also found out that students who start later do not start participating from the beginning units.

Some students who remain central over time have heavy postings on Unit discussion forums. Some of them retain their centrality over time due to heavy postings on other course forums. Also, most of these people belong to Cohorts 1 and 2 only. Another nice characteristic about these people is that they exhibit good forum behavior. Majority of these students who remained central started off as being answer persons, but most of them turned to discussion persons as time progressed. On the contrary, majority of the students who didn't remain central, started off as being discussion persons, but most of them turned to answer persons as time progressed. This re-emphasizes the importance of discussions in MOOC's, and its attached relevance to students' importance.

## 7.2 MOOC Discussions

Discussion in study groups becomes very sparse over time, though such study groups are meant to connect people having similar demographics like common work time(schedule), background, location etc. There can the following possible reasons for this: a)Limited knowledge of students, which is insufficient to spark off interesting discussions; b)The social phenomenon of Homophily or "birds of a same feather flocking together". This is where the course instructor can come in handy and introduce some interesting topics, or influential people from amongst those who post in these study groups can be helpful in igniting interest in these study groups.

Another interesting thing that we observe is when students usually return back to post back in old threads, there are there intense pockets of discussion. This motivates our clustering approach to group threads and distinguish between different conversation categories simultaneously going on in the MOOC. This knowledge will help course instructors to focus on a smaller subset of threads that need attention, instead of looking at the exponentially growing individual threads. Also, they can steer students towards more participation in course activities and engagement in discussions, by suggesting them the active and healthy threads.

## 7.3 Social network structure

The "Fantasy and Science Fiction: The Human Mind, Our Modern World" MOOC network consists of many small and dense cores throughout the network, instead of exhibiting a single core-periphery structure. Students are densely connected to different subparts of the social network graph. This happens due to multiple subgroup formation within the network because of thread starters and sub thread starters. Such subgroups are closely intra knit, but not well inter knit.

Apart from that, what we observe from the network skeleton or the bow tie structure is that, instructors should : a) Focus on the group of people in the 'Out' component. They should try to bring more people in the 'Out' component and help them to engage into the mainstream discussion going on in the course forums. For doing so, they may require to get support from students in the 'Scc', who drive these discussions; b)Furthermore, instructors should also motivate the sub thread starters to actively participate in discussions which they themselves initiate within threads, rather than merely acting as bridges between people. We notice that sub thread starters lie in a favorable structural position in the network, that is, they have higher betweenness centrality. So, once they are motivated enough, they can help disconnected groups of students to bond together with the discussion threads.

## 8. FUTURE RESEARCH DIRECTIONS

We plan to work in some closely related directions to dissect MOOC's further. Some interesting enhancements to our work will include comparative analysis of the currently studied beginner level MOOC course, with intermediate and advanced level courses to contrast the findings in forum participation and behavior. Adopting a semantic network perspective for analyzing MOOC's, so as to decipher the relation between words and discussion themes and find stylistic aspects of conversations that make specific groups of students distinctive in the networked learning community, would be fascinating to work on. Extraction of linguistic features in the conversations of special groups of students like cohorts, people with increasing and decreasing centrality can corroborate the connections of these individuals in the MOOC. Mapping these localized patterns to the linguistic patterns and topic models of the whole group of participating students in the MOOC, can help us to see what aspects of conversations are different and make these students distinct. This could be used in providing weekly reports on student discussion trends to the instructors.

Next, schemes for efficient labeling of every post/comment in all the forums using the structured Unit discussion forum need to be devised. The adopted approach should be able to clearly depict students returning back to discuss on forums later in time. This will give an overall idea about the kind of discussions across all forums in the MOOC. Possible approaches could leverage the first and last post/comment in each unit, sort them to get certain disjoint interval ranges for the post/comment and label all the posts/comments in this interval by the unit number which has maximum postings or using a probability distribution.

Leveraging the social network discovered behind the MOOC to study Information diffusion models for the MOOC can help us track how students reach a consensus, given a specific problem setting. Guille et al. [28] highlights some of the interesting research directions in this regard. Furthermore, both e-learning and the development of social network analysis can be benefited, if efficient mechanisms can be designed to balance the tradeoff between resource utilization (time, money etc.) and seeding of multiple cores that are evident in the network structure, so as to effectively communicate interesting ideas.

Examining how pre course enrollment motivations of students maps to their exhibited forum behavior in the MOOC is a closely related work that can follow up our current work. A preliminary analysis of this nature was performed by Belanger et al. [26]. However, the distinction that, because people had specific kind of prior motivations and therefore they behaved in fundamentally distinct ways in their conversations, was not brought out.

## 9. CONCLUSION

In this work, we demonstrated how social network analysis can help us in understanding informal connections between students participating in the MOOC. Through an application of computational social network methods having well established theoretical underpinnings, we deciphered the structural position of different student groups in the MOOC and its influence on their discussion behavior. We unveiled the participation patterns of active discussion initiators in the MOOC, while identifying bottlenecks to effective knowledge dissemination in the MOOC network. Then, to identify contrasting conversation categories simultaneously going on in the MOOC and enable instructors to focus on a important subset of discussion threads, we visualized the discussion forums over time, narrowing down from generic to more specific "Unit Discussion" forum. The fundamental motivation behind all the performed analysis was to facilitate the development of better and more accurate computer mediated support for students in the MOOC.

## 10. ACKNOWLEDGMENTS

This research work was carried out as a part of Summer Internship from June-August 2013 by the author, under the mentorship of Dr. Carolyn Penstein Rose, Language Technologies Institute, Carnegie Mellon University (CMU). The author would also like to acknowledge the support of Mr. David Adamson, PhD candidate, CMU for preparation of Coursera dataset, and his helpful insights on the problem. One of the outcomes of this work related to dropout analysis has been published at NIPS workshop on Data Driven Education 2013.